\documentclass[11pt]{article}

\usepackage{times}
\usepackage{mathtools}
\usepackage{graphicx}
\usepackage{epsfig}
\usepackage{epstopdf}
\usepackage{amsmath}
\usepackage{color}
\usepackage[normalem]{ulem}

\newenvironment{sciabstract}{%
\begin{quote} \bf}
{\end{quote}}

\newcounter{lastnote}

\title{Global Patterns of Synchronization in Human Communications}

\author
{Alfredo J. Morales $^{1,2*}$, Vaibhav Vavilala $^{1}$, Rosa M. Benito $^{2}$, Yaneer Bar-Yam $^{1}$ \\
\normalsize{$^{1}$ New England Complex Systems Institute (NECSI),}\\
\normalsize{210 Broadway St., Cambridge MA 02139, USA.}\\
\normalsize{$^{2}$ Grupo de Sistemas Complejos, Universidad Polit\'ecnica de Madrid,}\\
\normalsize{ETSI Agr\'onomos, Av. Computlense S/N, 28040, Madrid, Spain.}\\
\normalsize{$^\ast$ E-mail: alfredo@necsi.edu.}
}

\date{}

\begin{document}

\maketitle

\begin{sciabstract}

Social media are transforming global communication and coordination. The data derived from social media can reveal patterns of human behavior at all levels and scales of society. Using geolocated Twitter data, we have quantified collective behaviors across multiple scales, ranging from the commutes of individuals, to the daily pulse of 50 major urban areas and global patterns of human coordination. Human activity and mobility patterns manifest the synchrony required for contingency of actions between individuals. Urban areas show regular cycles of contraction and expansion that resembles heartbeats linked primarily to social rather than natural cycles. Business hours and circadian rhythms influence daily cycles of work, recreation, and sleep. Different urban areas have characteristic signatures of daily collective activities. The differences are consistent with a new emergent global synchrony that couples behavior in distant regions across the world. A globally synchronized peak that includes exchange of ideas and information across Europe, Africa, Asia and Australasia. We propose a dynamical model to explain the emergence of global synchrony in the context of increasing global communication and reproduce the observed behavior. The collective patterns we observe show how social interactions lead to interdependence of behavior manifest in the synchronization of communication. The creation and maintenance of temporally sensitive social relationships results in the emergence of complexity of the larger scale behavior of the social system.

\end{sciabstract}

\section{Introduction}
The functioning of complex systems, like human societies or living organisms, depends not only upon the individual functionalities of their parts but upon the coordination of their actions. 
{\color{black} Self-sustaining activities, such as economic transactions and associated communications, occur through interactions among people, creating dependencies among their actions.}
A central challenge for both sociology and economics is our ability to characterize the collective action of individuals that together become the aggregate activity that comprises our society \cite{bettencourt2013origins}.
Recent studies have shown that these processes can be observed by looking at communication patterns among individuals in social groups \cite{pentland2014social}.
Here we analyze Twitter data to describe the underlying dynamics of social systems. In particular, we study collective activities across geographical scales, from areas smaller than one square kilometer up to the global scale.

The recent explosion of social media is radically changing the way information is shared among people and therefore the properties of our society. These new mechanisms allow people to easily interact with each other and to affordably exchange and propagate pieces of information at multiple scales. As a consequence, people may be able to engage in types of complex tasks previously dominated by organizations structured for a particular purpose \cite{hbr2015}. 
By looking for patterns in the aggregate data, we can retrieve structural and dynamical information about the social system \cite{Lazer2009}. This represents an unprecedented opportunity to study social systems across many scales.
Traditional surveys of small samples, which are typically limited to a few questions, do not have the scale and frequency to capture such population dynamics  {\color{black} \cite{Lazer2009}.}

\begin{figure}[t]
\includegraphics[width=\textwidth]{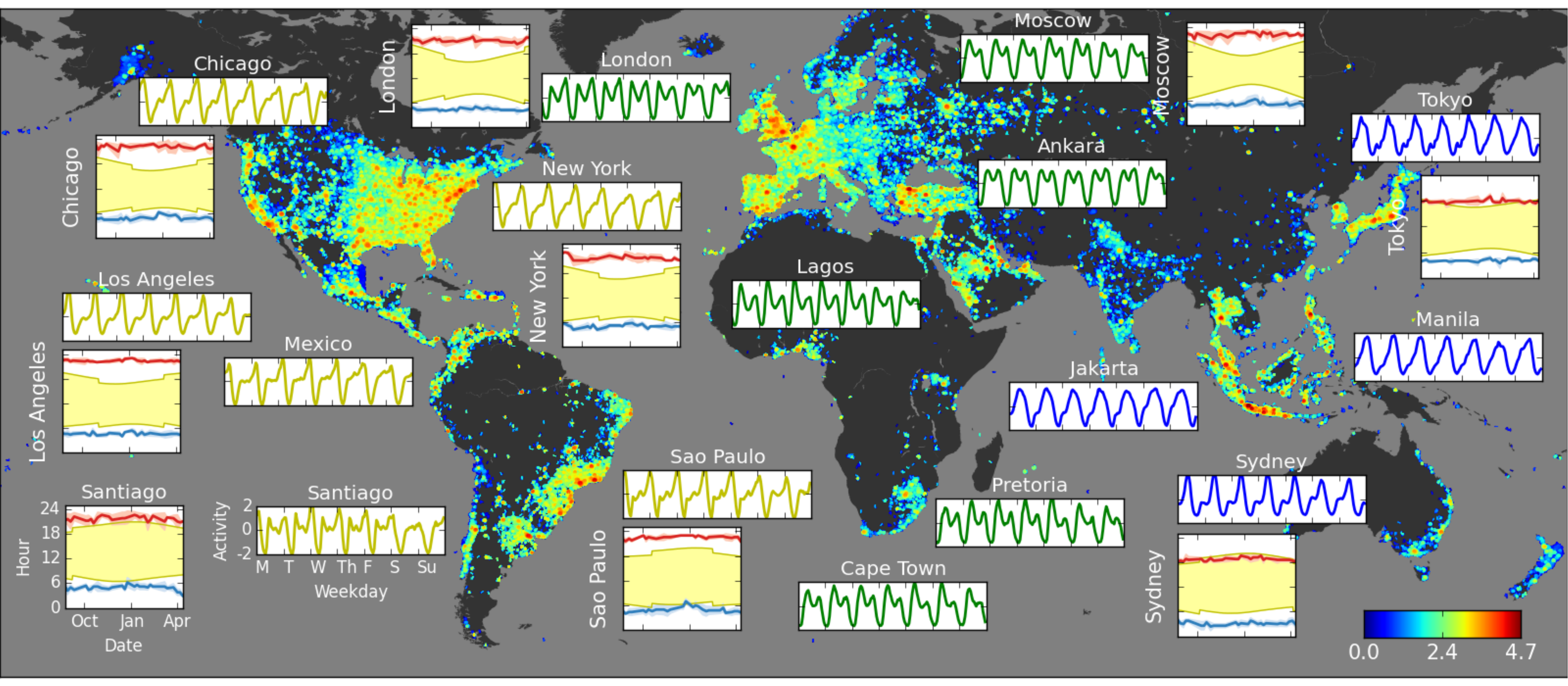}
\caption{Global Twitter Activity. Background map: Twitter activity in each 0.25$^\circ$x0.25$^\circ$ geographic area (base-10 log scale  at lower right). Rectangular insets: Average week of Twitter activity of selected cities in Universal Time (UTC) after subtracting the mean and normalizing by the standard deviation. Square insets: Low ({\color{black}blue}) and high ({\color{black}red}) points of Twitter activity of several urban areas compared to daily sunlight periods (yellow) during the nine month observation period (scales on lower left shown for Santiago are the same for all cities). {\color{black} An animated visualization of the global Twitter activity is shown in the supplementary electronic material (video S1). }}
\label{fig:ActivityMap}
\end{figure}

As highly concentrated social systems, cities are manifestly complex systems with emergent properties \cite{batty2008size}. They are self-organized entities made up of multiple complex agents that engage in larger-scale, complex tasks. Moreover, cities have multiscale structures individually through fractal growth and collectively through size distributions. Their structural patterns have been modeled by scaling laws \cite{bettencourt2013origins}, archetypes of streets layouts \cite{buhl2006topological,louf2014typology} and land use \cite{decraene2013emergence}. Human generated data has been used to understand the dynamical behavior of inhabitants and their impact on the city functioning. Patterns of human activity and mobility reveal the spatial structure of collective interactions \cite{ratti2006mobile,louail2014mobile} and the dynamical properties of urban functional areas \cite{toole2012inferring, frias2012characterizing}. Many of these studies use mobile phone data which is available only for a small set of cities. 


In this work, we analyze over 500 million geolocated tweets, posted between August 1st, 2013, and April 30th, 2014, to explore patterns of social dynamics in urban areas around the world. We collected these data using the Twitter streaming Application Programming Interface (API) \cite{twitapi2015}, which provides over 90\% of the publicly available geolocated tweets \cite{morstatter2013sample} in real time. Twitter is an online social network whose users share ``micro-blog" posts from smartphones and other personal computers. 
Its population trends younger, wealthier and urban \cite{duggan2013demographics,Mislove2011}, which makes it a good probe of the dynamics of young workers in cities. Geolocated tweets 
provide a precise location of the {\color{black} individuals that post messages}, and represent around 3\% of the overall {\color{black} Twitter stream} \cite{FM4366}. Twitter activity has been analyzed to understand human sentiments \cite{Golder2011}, news sharing networks \cite{Herdagdelen2012} and influence dynamics \cite{Morales2014},
as well as global patterns of human mobility \cite{10.1371/journal.pone.0105407}, activity \cite{Lenormand20150473} and languages \cite{10.1371/journal.pone.0061981}.

\begin{figure}
\begin{center}
\includegraphics[width=6in]{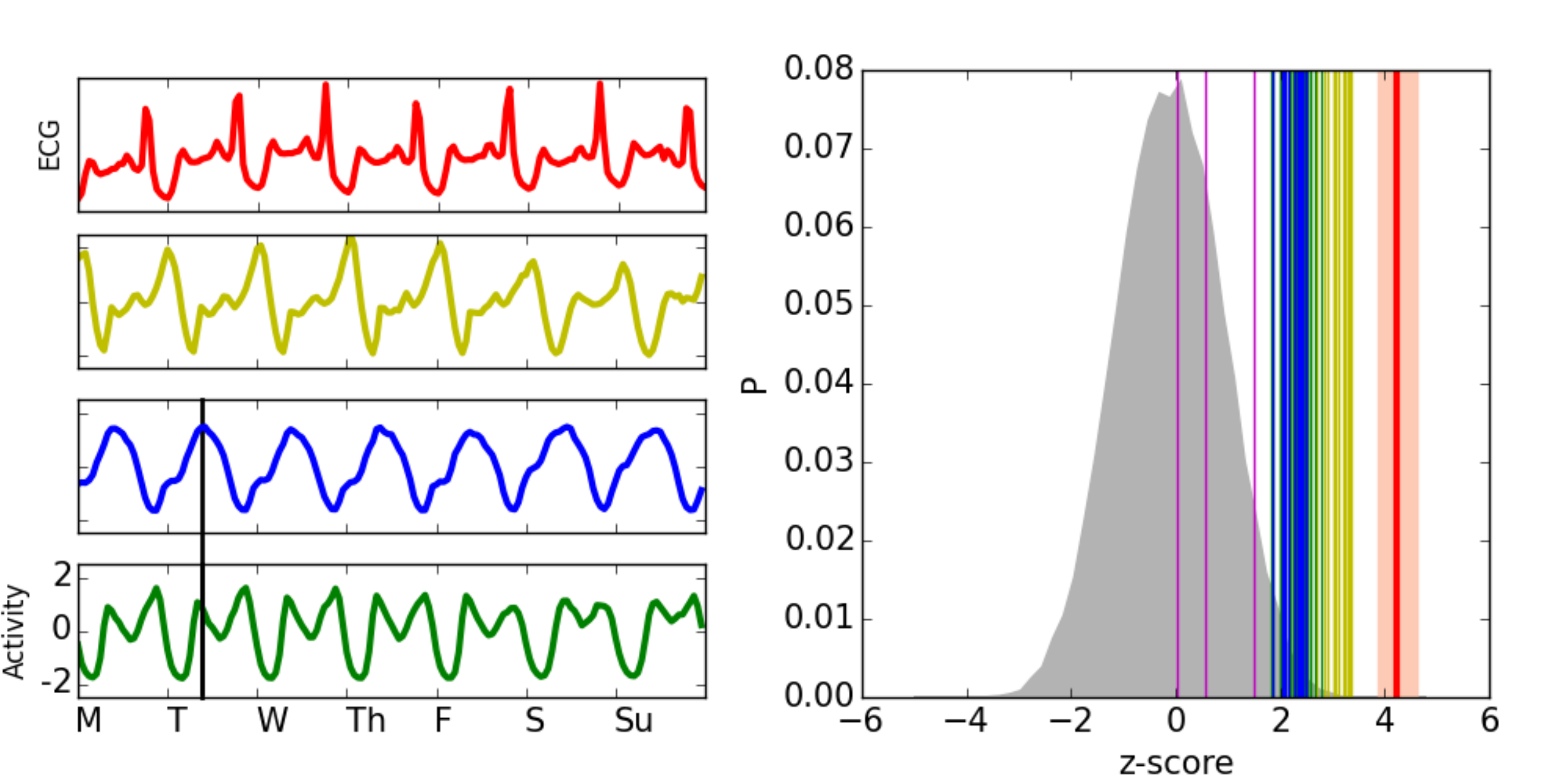}
\caption{ Correlation of the temporal dynamics of cities and heart beats. Left panels: An example of heart beat ECG signal at approximately 80 beats per minute (red) and the average week of activity for three cities: S\~ao Paulo (yellow), Jakarta (blue) and London (green). Vertical black lines show the time of synchronization (see text). Right panel: Correlation of the heartbeat with 50000 random series (gray curve), other heartbeats (red line), periodic signals (magenta lines, from left to right: sawtooth, squared and sinusoid), and all urban areas colored by group determined by clustering analysis (see Supplement). }\label{fig:HeartBeatCorr}
\end{center}
\end{figure}

\section{Urban Activity}

In Figure \ref{fig:ActivityMap}, we show the hourly number of tweets during an average week for a few major metropolitan areas (rectangular insets) on top of a map representing the global density of tweets during an average day (52 metropolitan areas across the world are in the Supplement). 

{\color{black}

To construct the average week, we count the number of tweets in each hour in the urban area $i$ during the observation period ($N$ weeks), $s_{i,t}$. Time can be written in terms of the hour of the week $t=t' \mod W$, where $t \epsilon \{1,...,W\}$, $W=168$ is the number of hours per week and $t'$ is the number of hours since the start of the observation period. We then average the corresponding hours of each week (with the same value of $t$) after normalizing by the standard deviation for that day: 

\begin{equation}
\label{eq:week}
s’_{i,t} = \frac{1}{N} \sum_{w=0}^{N-1} \left( \frac{ s_{i,t + W(w-1)} - \frac{1}{D}\sum\limits_{\sigma=0}^{23} s_{i,\sigma+ D\lfloor{t/D}\rfloor + W(w-1)}}
{ \sqrt{\frac{1}{D}\sum\limits_{\sigma=0}^{23} \left(s_{i,\sigma+ D\lfloor{t/D}\rfloor + W(w-1)} - \frac{1}{D}\sum\limits_{\sigma=0}^{23} s_{i,\sigma+ D\lfloor{t/D}\rfloor + W(w-1)}\right)^2}} \right) 
\end{equation}

\noindent where we use a 24 hour daily index $\sigma \epsilon \{0,...,D-1\}$, $D=24$, to define the daily average and standard deviation.  $\lfloor{t/D}\rfloor$ is the integer part of $x$. We further define an average day as the average over corresponding hours of the days of the average week: 

\begin{equation} 
\label{eq:day}
\hat{s}_{i,\sigma}= \frac{1}{\Omega}\sum_{d=0}^{\Omega-1} s_{i,\sigma + D d}
\end{equation}

\noindent where $\sigma \in \{0,\ldots,D-1\}$ and  $\Omega=7$.

 }

\begin{figure}
\begin{center}
\includegraphics[width=6in]{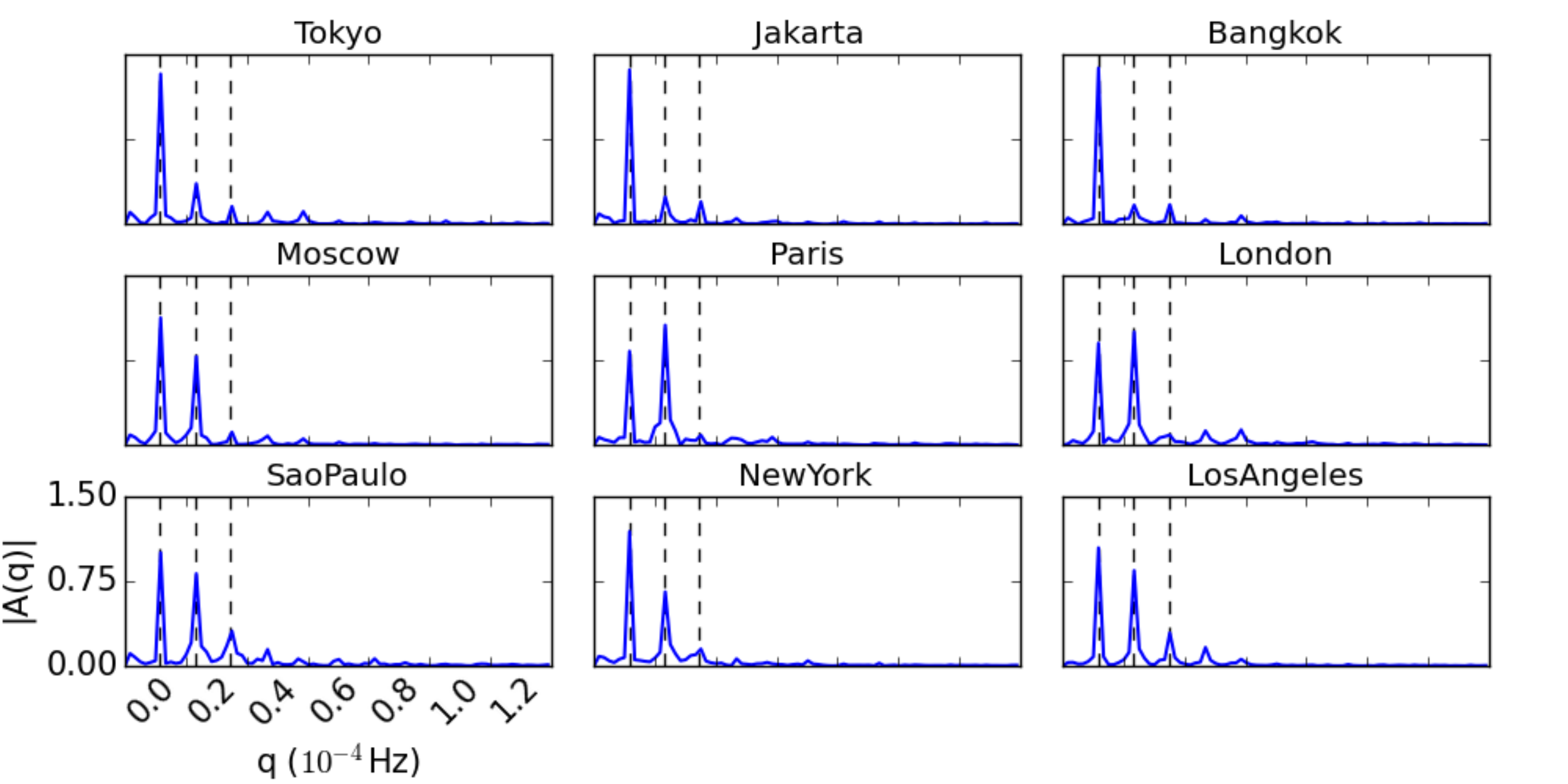}
\caption{Spectral analysis of Twitter activity (amplitude of the Fourier Transform) of major urban areas (more cities in the Supplement). The dashed lines indicate (from left to right) the frequencies (q) corresponding to the periods of 24 hours, 12 hours and 8 hours respectively.}\label{fig:fourier}
\end{center}
\end{figure}

\begin{figure}
\begin{center}
\includegraphics[width=6in]{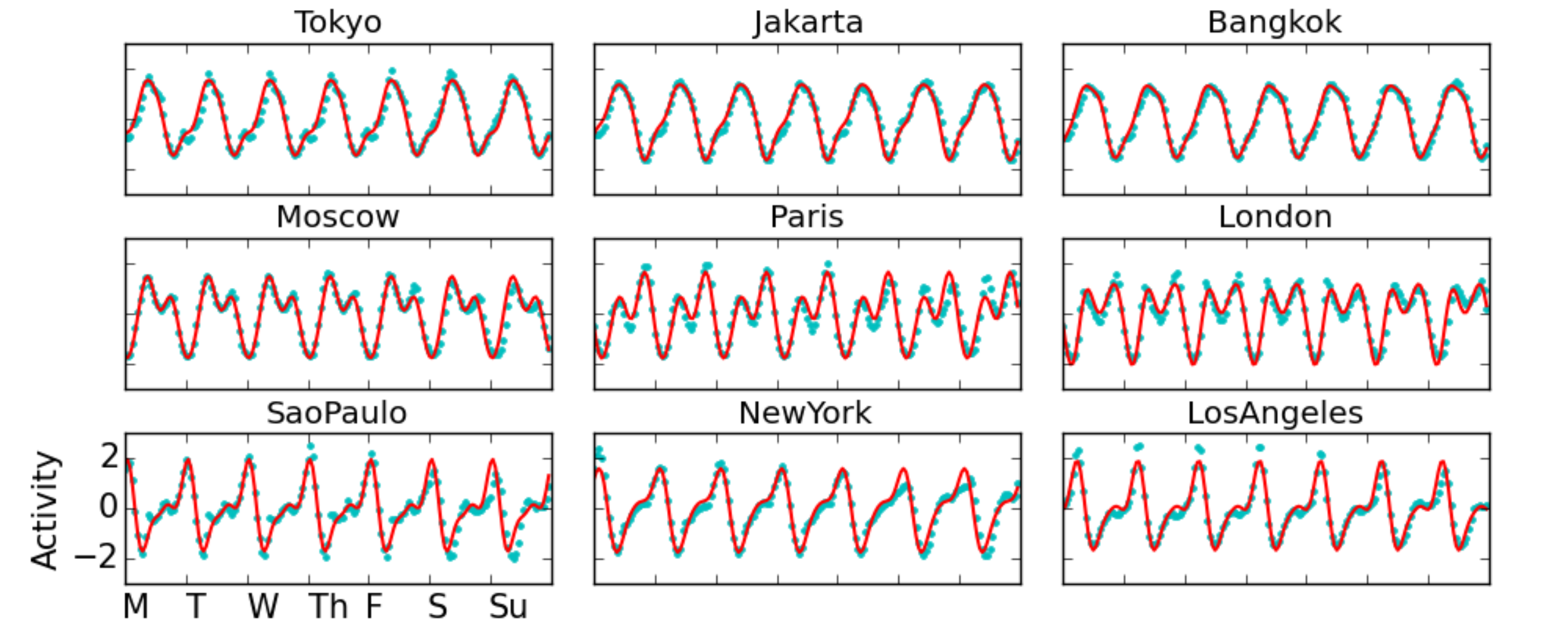}
\caption{Modeling the Twitter activity of major urban areas by spectral decomposition (more cities in the Supplement). Hourly number of tweets during an average week (blue dots) are compared with model results from Eq. \ref{eq:model_series} (red curves).}\label{fig:model}
\end{center}
\end{figure}

Overall, cities present similar patterns of collective behavior, cycling between peaks and valleys of activity. Such patterns are also found in phone calls \cite{2008JPhA...41v4015C}, electricity consumption \cite{Pielow2012533} and emails \cite{wang2011information}. Peaks occur during daytime or evening, indicating that people are awake and active, simultaneously tweeting from work, recreation or residential areas, whereas valleys occur during night and sleeping hours, indicating that people are resting and inactive. 

The time series' regular behavior indicates that people synchronize their activities throughout the day. 
This synchrony is not solely due to external factors like light and dark or due to biological factors like circadian rhythms \cite{opac-b1117004}. 
The second set of insets show 
low (blue) and high (red) points of the activity along with the time of sunrise and sunset over the year (yellow shadowed area). The wide range of light and dark times does not cause a comparable shift in activity times. 
{\color{black} We calculated the time difference between the series' morning valleys and sunrise times, as well as the time difference between the series' afternoon peaks and sunset times. We grouped these time differences in 10-day intervals, calculated their distributions and compared with a null hypothesis that the distributions do not change. We found that the average largest and shortest time differences over the observation period are significantly different ($p<0.01$ for equatorial cities and $p<0.001$ otherwise), indicating that variations in the sunset or sunrise times do not determine the times of peaks or valleys of activity.}

{\color{black}Our economic system is based upon transactions, communications and coordination involving the activities of multiple workers.
The completion of tasks within a given time frame depends upon the joint availability of workers either simultaneously or in the correct sequence \cite{van2004workflow}. 
Synchrony has its costs, as commuting traffic jams illustrate, but many activities are less effective or impossible to do without it. Synchrony enables couples in relationships to share waking and sleeping schedules. Broad categories of coordinated behaviors like work, leisure and sleep make up the temporal superstructure which organizes all other tasks within the daily timeline of a city. 
For workers on a 9-5 schedule, working at the same time every day enables them to meet to conduct business activities together, whether in person or by telephone. Others who work outside of regular business hours are able to provide services like entertainment and shopping opportunities to those who work during the primary shift. 
The coordination of social activities during these off hours is possible because of the synchrony of standard work and rest hours. Biological circadian rhythms are important for the synchronization of sleeping hours \cite{opac-b1117004} and the interactions of couples contribute to the synchronization of activities.}

Similar patterns arise in the biological activity of living organisms, like heartbeats or respiration, or in their collective activity, like in termite colonies \cite{Jamali20111471} or ecosystems like forests \cite{Grace03111995}. 
Heartbeats in particular have properties that appear in some ways similar to urban dynamics. For comparison, we show in the left panel of Figure \ref{fig:HeartBeatCorr} the electrocardiogram (ECG) of a 43-year-old male \cite{Greenwald1986,al2000physiobank} together with the Twitter activity from S\~ao Paulo (yellow), Jakarta (blue) and London (green). 
{\color{black} The similarity between heartbeats and Twitter activity is remarkable and quantified in the right panel of Fig. \ref{fig:HeartBeatCorr} by correlating ECG signals of individual heartbeats with urban Twitter activity (also see Supplement B). The signature of a regular heartbeat is obtained by segmenting the ECG into equal average heartbeat intervals that are further divided into 24 segments (similar to the 24 hours in a day). This signature is correlated with the ECG and Twitter activity. The correlation is done by placing the minimum value in each period for both signals at the center of the correlation window. The heartbeats are highly correlated with the Twitter activity in urban areas, just less than the correlation of heartbeats with each other. The correlations with the time series are compared with those of 50,000 random time series, as well as with those of periodic series, in the right panel of Fig. \ref{fig:HeartBeatCorr}.}
While the high level of correspondence is not essential to our discussion, the reasons for it can be understood. Both regular heart activity and human urban activity have three primary periods. The heart experiences a strong (ventricular) contraction, a secondary (atrial) contraction and a period in which both are relaxing. Human urban collective activity has a primary work shift, a secondary work and recreation shift, and a sleep shift. 

The three fold cyclical Twitter behavior is further analyzed by spectral decomposition (Fourier transform). The spectral behavior of the Twitter activity series from 6 major cities are shown in Fig. \ref{fig:fourier} and the remaining cities are shown in the Supplement. All frequency spectra have three significant components at 24h, 12h and 8h (dashed lines). The first is due to variations associated to the daily cycle, the second to variations during 12 hours periods, night and day, the third corresponds to periodic variations within work, recreation and sleep `shifts.' 

Based on the frequency decomposition, we can model the kinetics by adding three sinusoid signals of 24, 12 and 8 hours period respectively. In the model, the activity $s(t)$ is defined as:

\begin{equation}
\label{eq:model_series}
s(t)=a_{24} \sin(\frac{2\pi t}{24}+\theta_{24})+a_{12} \sin(\frac{2\pi t}{12}+\theta_{12})+a_{8} \sin(\frac{2\pi t}{8}+\theta_8)
\end{equation}

\noindent where $t$ is time in hourly resolution, $\theta$ represents the respective signal phase, and $a$ is the signal amplitude in the range [0,1].
We fit the parameters  $\theta$ and $a$ for each time series by minimizing the quadratic error. The model fits well the observed data ($p<0.001$), as shown in Figure \ref{fig:model} and Supplement F.

\begin{figure}
\begin{center}
\includegraphics[width=\textwidth]{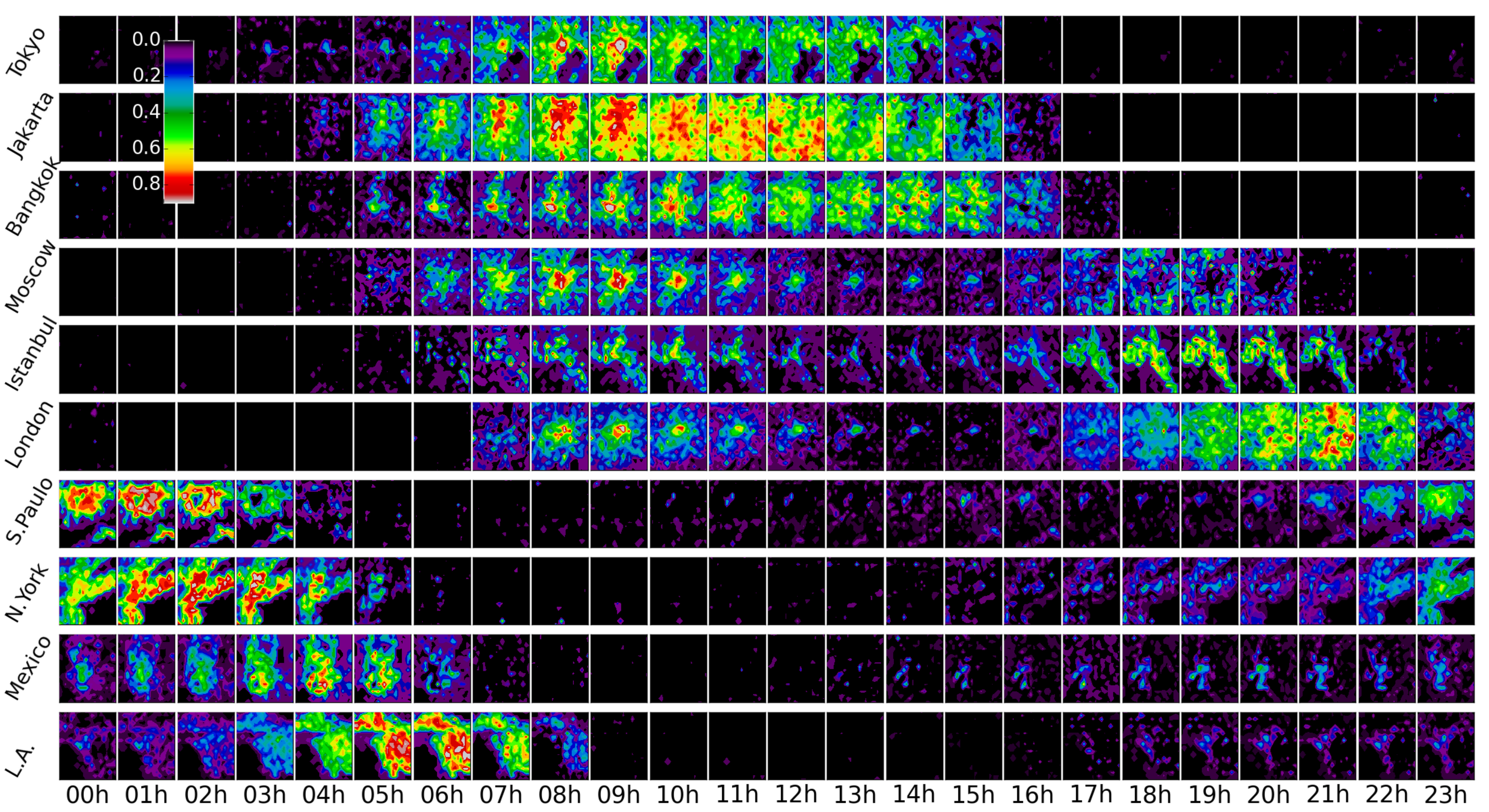}
\caption{Spatio-temporal dynamics of Twitter activity in urban areas. Each row shows activity during an average day according to UTC time for the specified city. Colors indicate the normalized excess of activity from the average value at that location (scale shown in figure). {\color{black} Animated visualizations of Twitter activity in urban areas are shown in the supplementary electronic material (videos S2-4). }}\label{fig:UrbanNebulae}
\end{center}
\end{figure}

\section{Spatial Patterns}

\begin{figure}[t]
\begin{center}
\includegraphics[width=6in]{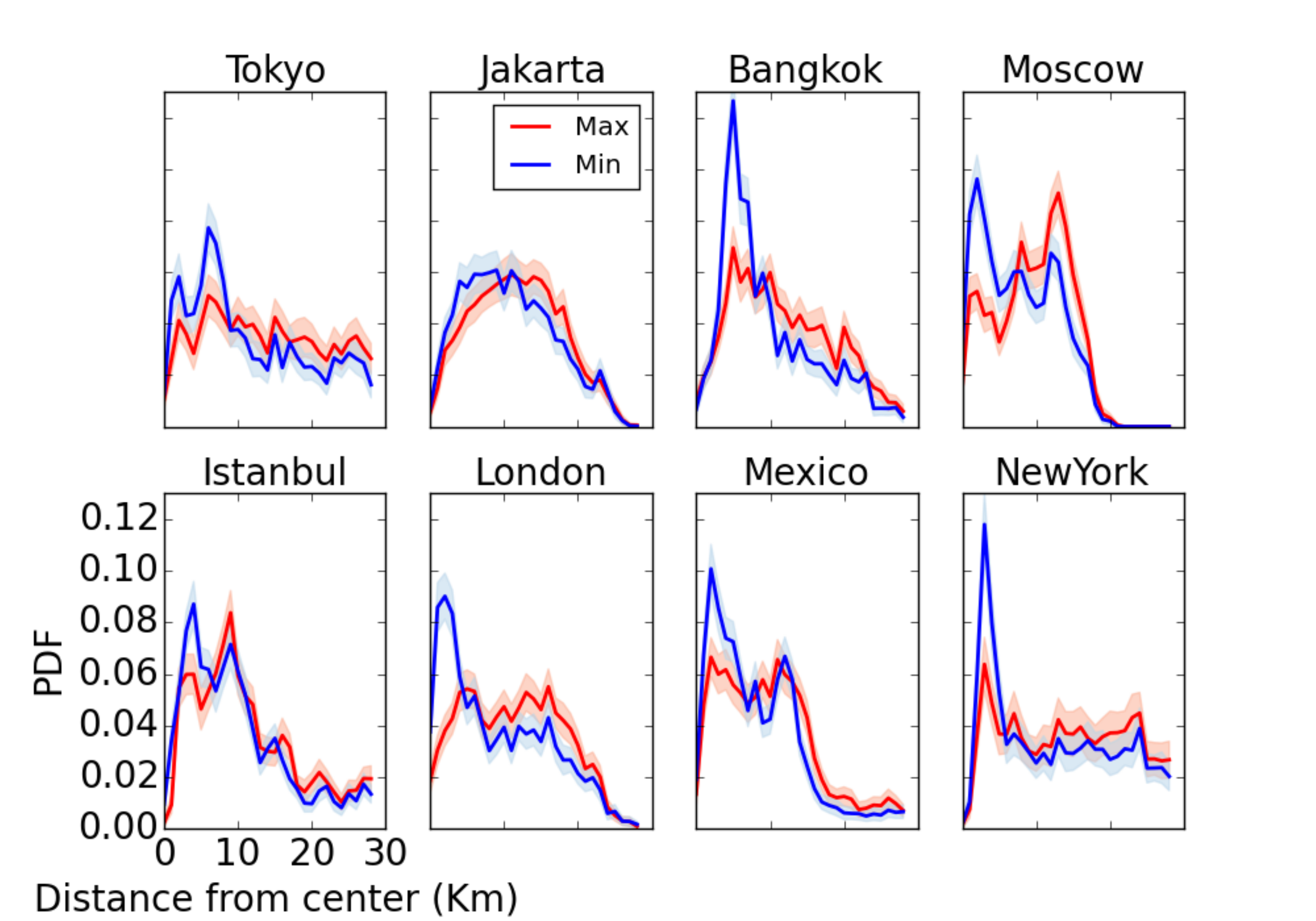}
\caption{Distribution of tweets as a function of the distance to the city center (calculated as the center of gravity of the spatial activity of each city) during the most contracted (blue) and expanded (red) periods between 9 am and midnight in major urban areas. }\label{fig:distanceCenter}
\end{center}
\end{figure}

{\color{black} While there are social behaviors that differ among individuals and cultures, there are also those that are common and universal such as the very existence of cities and urban areas}. 
{\color{black}In cities, many} people concentrate in a few but dense business or commercial areas during work hours, whereas they disperse to typically sparse residential areas at rest hours. 
We expose this behavior by looking at both spatial variation of Twitter activity and individual mobility patterns. 
{\color{black} We first disaggregated the average day of Twitter activity into a lattice of 20x20 patches in each urban area. In Figure \ref{fig:UrbanNebulae}, we show patches of local activity for 10 major metropolitan areas after subtracting the average, normalizing by the standard deviation and coloring the activity above average values (see Eq. \ref{eq:day} where $i$ represents each patch).  The average and standard deviation are for each patch, rather than the whole city (see Supplement C)}. 

\begin{figure}[t]
\begin{center}
\includegraphics[width=6in]{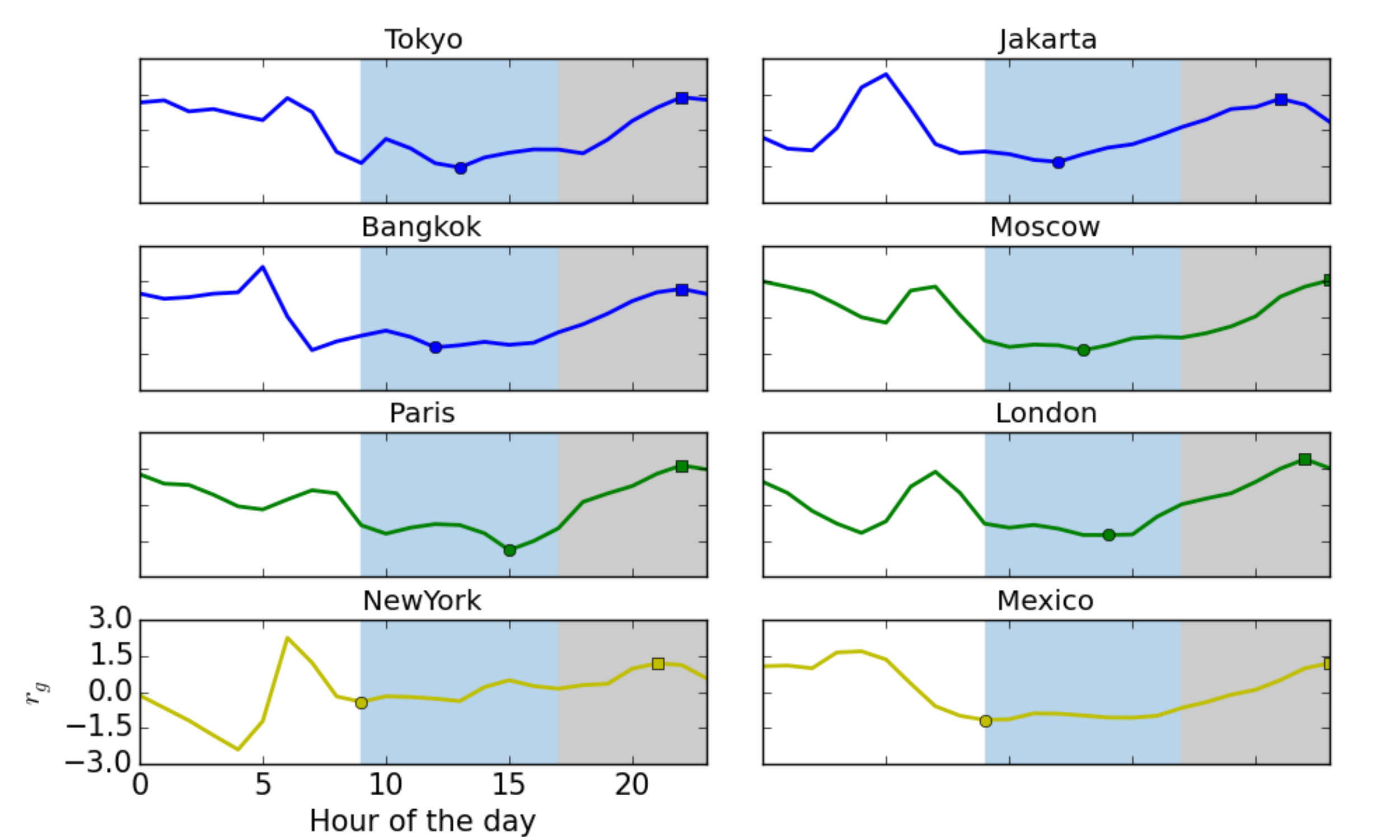}
\caption{Radius of gyration ($r_g$) of tweets during an average day in major metropolitan areas. Series are shown in local time and have been rescaled by subtracting the average and normalizing by the standard deviation. Blue shading indicates standard work hours (9-5) and gray shading indicates standard recreation and rest hours before midnight.}\label{fig:radius}
\end{center}
\end{figure}

In all cities, there are peaks of intense activity occurring near central areas which expand towards more peripheral areas over time. 
{\color{black}  In Figure \ref{fig:distanceCenter}, we show the distribution of tweets as a function of the distance to the city center, during the most contracted (blue) and expanded (red) times between 9 am and midnight. The city center is calculated as the center of gravity of the spatial activity. For each city, we compared these two distributions, with a null hypothesis that they are similar to each other. After applying bootstrapping, we found that the average distance of tweets to city center differs significantly during the most contracted and expanded times ($p<0.001$). This effect is also manifest in the hourly variation of the radius of gyration ($r_g$) in Figure \ref{fig:radius}. We calculate the hourly radius of gyration by performing bootstrapping at each hour of the day across the whole observation period and averaging across the same hour of all days. The radius of gyration of each city shows a daily cycle of contraction and expansion during work (blue shaded region in Fig. \ref{fig:radius}) and rest hours (gray shadowed regions in Fig. \ref{fig:radius}) respectively. Table \ref{tab:radius} shows the maximum and minimum values of the radius of gyration between 9 am and midnight. In all urban areas, the radius of gyration varies significantly ($p<0.001$).

We apply bootstrapping in the following way. For each hour of the day $h$, we take $N=500$ sample sets $\phi_{h,i}$ of $M=500$ tweets $j$ and calculate the tweets distance to the city center $d_j$. Then we calculate the average distance $\mu_{\phi_{h,i}}=1/M\sum_{j \in \phi_{h,i}}{d_j}$ for each sample set, as well as the average distance across all sample sets $ \hat{\mu}_h =1/N\sum_i {\mu_{\phi_{h,i}}}$, standard deviation $\hat{\sigma}_h=\sqrt{1/N\sum_i{(\hat{\mu}_h-\mu_{\phi_{h,i}})^2}}$ and standard error  $\bar{\sigma}_h=\hat{\sigma}_h/\sqrt{M}$. We determine $h_{max}$ and $h_{min}$ as the times where $\hat{\mu}_h$ is either maximum of minimum between 9am and midnight. We determine the p-value that $ \hat{\mu}_{h_{max}}$ and $\hat{\mu}_{h_{min}}$ are part of the same distribution. Since $ \hat{\mu}_{h_{max}}$ and $\hat{\mu}_{h_{min}}$ are selected from 24 values for extremality, we used extreme value theory \cite{opac-b1126783}. Using Monte Carlo sampling, the p-value that $\hat{\mu}_{h_{max}}$ is within the $ \hat{\mu}_{h_{min}} $ distribution and vice-versa is $p<0.001$ in all cases shown and over 90\% of the cities present in the supplement. Analogous procedures are performed with the radius of gyration obtaining similar results.

}

\begin{table}
\caption{Maximum and minimum radius of gyration ($r_g$) of tweets during an average day for 8 major metropolitan areas in kilometers. The difference of their magnitudes is significant ($p<0.001$). Other cities are in Supplement C.}
\label{tab:radius}
\begin{center}
\begin{tabular}{ |c | c | c || c | c | c| }
 \hline
   City & Max. $r_g$ &Min. $r_g$ & City &Max. $r_g$ & Min. $r_g$ \\ \hline
 Tokyo & 23.855 $\pm$  0.552  &  21.413  $\pm$  0.612   &   Jakarta & 13.671 $\pm$  0.265  &  12.826  $\pm$  0.270  \\ \hline
 Bangkok & 15.047 $\pm$  0.382  &  13.251  $\pm$  0.401   &   Moscow & 11.157 $\pm$  0.183  &  9.516  $\pm$  0.219  \\ \hline
 Paris & 4.066 $\pm$  0.053  &  3.690  $\pm$  0.061   &   London & 13.545 $\pm$  0.255  &  11.892  $\pm$  0.320  \\ \hline
 New York City & 38.150 $\pm$  1.119  &  35.572  $\pm$  1.158   &   Mexico & 12.977 $\pm$  0.368  &  11.135  $\pm$  0.359  \\ \hline

\end{tabular}
\end{center}
\end{table}

\begin{figure}
\begin{center}
\includegraphics[width=6in]{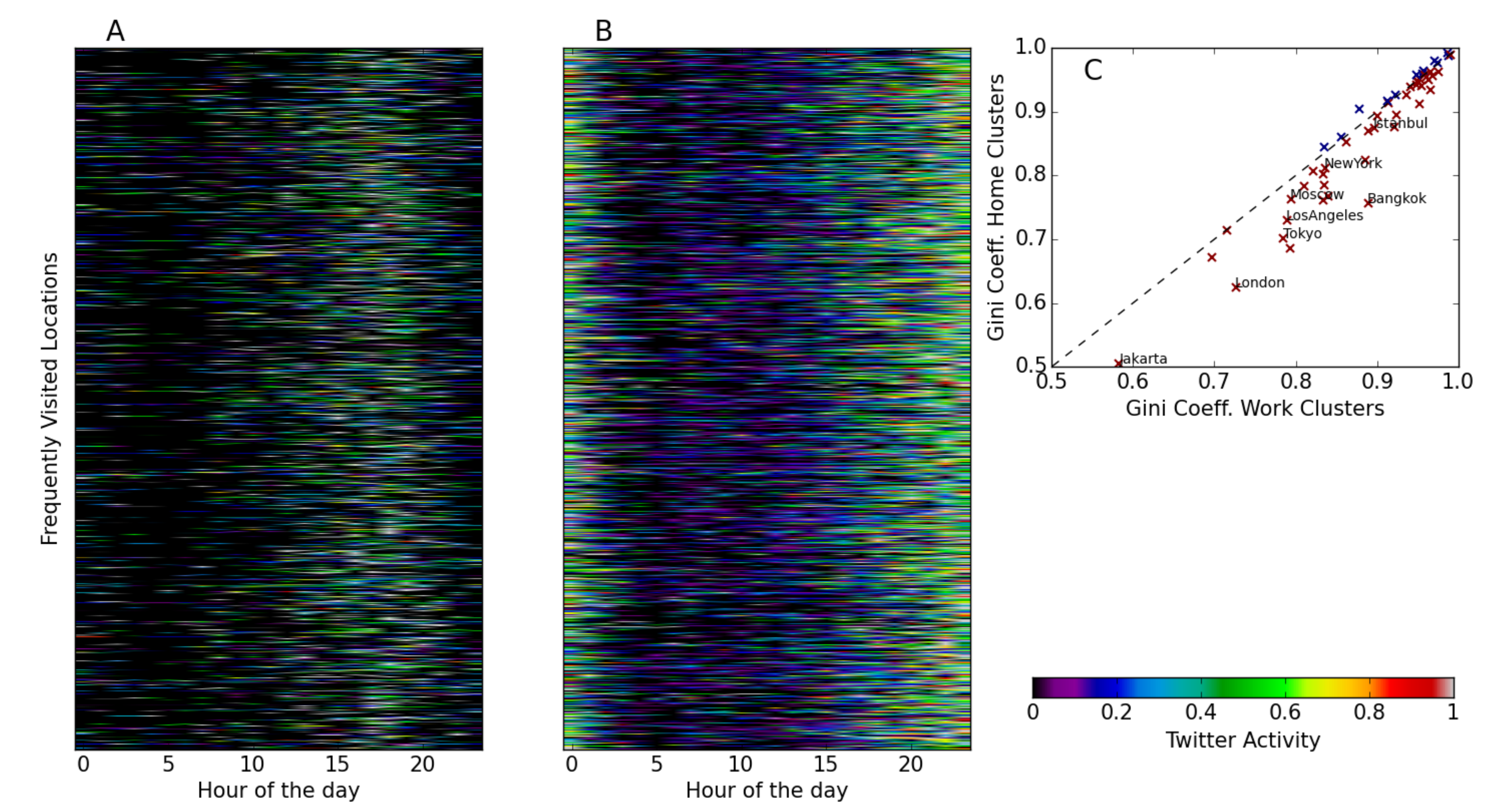}
\caption{Clusters of frequently visited locations according to the time of the day of individual activity. Dominant location clusters have primary activity during either conventional work (9-5) or rest hours according to local time. A. Total activity in all cities at clusters whose primary activity is during conventional work hours. B. Like A but for clusters whose primary activity is not during work hours (scale is shown in the figure). C. Geographical heterogeneity (similar to Gini) coefficient of conventional work and rest hours locations.
}\label{fig:mobility_supp}
\end{center}
\end{figure}

\begin{figure}
\begin{center}
\includegraphics[width=6in]{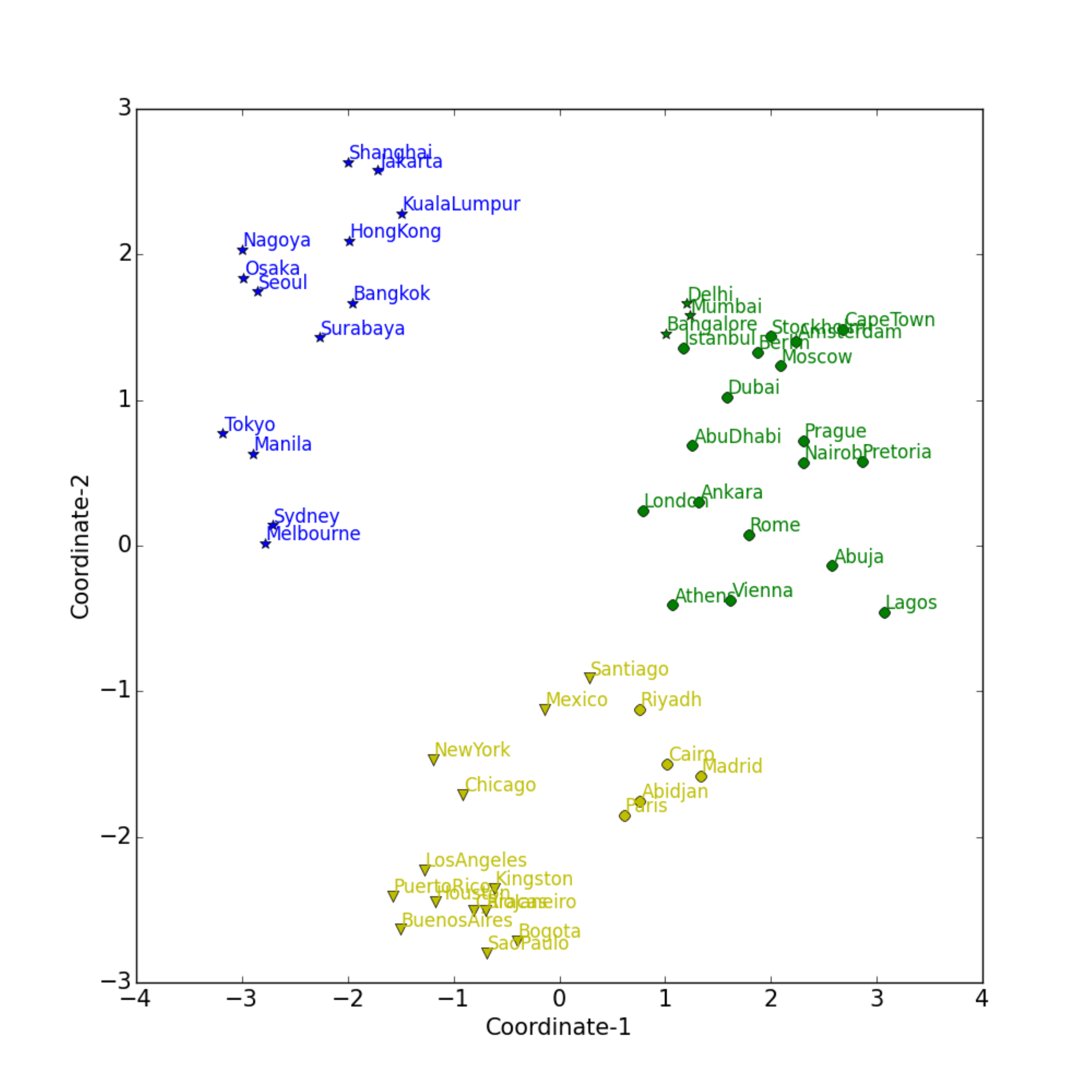}
\caption{Multidimensional scaling and clustering of city activity according to their time series vectors. Colors indicate the results of the clustering algorithm (see text). Star symbols represent Asian and Oceanian cities. Circular symbols represent Mideastern, European and African cities. Triangular symbols represent South and North American cities. Axes correspond to reduced dimensions obtained from multidimensional-scaling.}\label{fig:MDSClustering}
\end{center}
\end{figure}

Second, we analyzed the most frequently visited locations of each user. By applying a clustering algorithm \cite{macqueen1967}, two clusters of frequently visited locations were identified (see Fig. \ref{fig:mobility_supp}). 
In one cluster, most tweets are sent during work hours (after 9am and before 5pm), while in the other cluster most tweets are sent during rest hours (after 8pm and before 2am).
This technique has been  
used to identify work and home locations from mobile phone data \cite{Becker:2013:HMC:2398356.2398375}. 
In general, work locations are less homogeneously distributed than residential locations.
{\color{black}We counted the number of work and residential locations in each patch of the 20x20 grids respectively. We found that work locations are more spatially clustered, which results in a higher spatial heterogeneity (similar to Gini) coefficient (Fig. \ref{fig:mobility_supp} C).}
{\color{black} Moreover, the average distance between residential locations with respect to the city center is significantly larger from that of work locations ($p < 0.001$ for half of the analyzed cities), indicating that home locations are more widespread.}

Despite 
overall similarities, 
cities have distinct signatures in the number and shape of peaks of activity both in time (Fig. \ref{fig:ActivityMap} and \ref{fig:HeartBeatCorr}) and space (Fig. \ref{fig:UrbanNebulae}). 
 {\color{black} By applying a clustering algorithm \cite{macqueen1967} to the time series vectors, starting from the minimum of activity, we found an optimal partition \cite{ROUSSEEUW198753} of three clusters of cities with remarkable cultural and regional affinity. These clusters have been colored accordingly in Figures \ref{fig:ActivityMap} and \ref{fig:HeartBeatCorr}, as well as in the dimensionally reduced space \cite{borg2005modern} shown in Figure \ref{fig:MDSClustering}. 
One class of cities, including Jakarta and other Asian and Oceanian cities (blue series in Fig. \ref{fig:ActivityMap} and blue symbols in Fig. \ref{fig:MDSClustering}), has a single large peak of activity during the day. Another class, including S\~ao Paulo and multiple North and South American cities (yellow series in Fig. \ref{fig:ActivityMap} and yellow symbols in Fig. \ref{fig:MDSClustering}), has two small peaks of activity in the morning and a large peak in the evening. Finally, a third class, including London and multiple Mideastern, European and African cities (green series in Fig. \ref{fig:ActivityMap} and green symbols in Fig. \ref{fig:MDSClustering}), has two equally sized peaks of activity, respectively at morning and afternoon. }Differences are also manifest spatially (Fig. \ref{fig:UrbanNebulae}). Asian cities (top three rows) gradually increase their activity (colored patches), showing spatial peaks, and rapidly decrease (black patches). European cities (middle three rows) have a strong spatial peak of activity in the morning near the center of the city and other dispersed peaks in the afternoon at peripheral areas. Finally, North and South American cities (bottom three rows) have several smaller peaks of activity around multiple centers and one large peak at the end of the day throughout the city. Interestingly, the morning peaks of the European cities coincide in time with the large afternoon peaks of Asian cities (see columns 8h to 10h), which is an indication of synchrony.
This synchrony is manifested as simultaneous peaks of activity in the time series (vertical black lines in Fig. \ref{fig:HeartBeatCorr} and Supplement A).
 It turns out that some of these differences can be traced to global patterns of behavior.

\section{Global Synchrony}

\begin{figure}
\begin{center}
\includegraphics[width=6in]{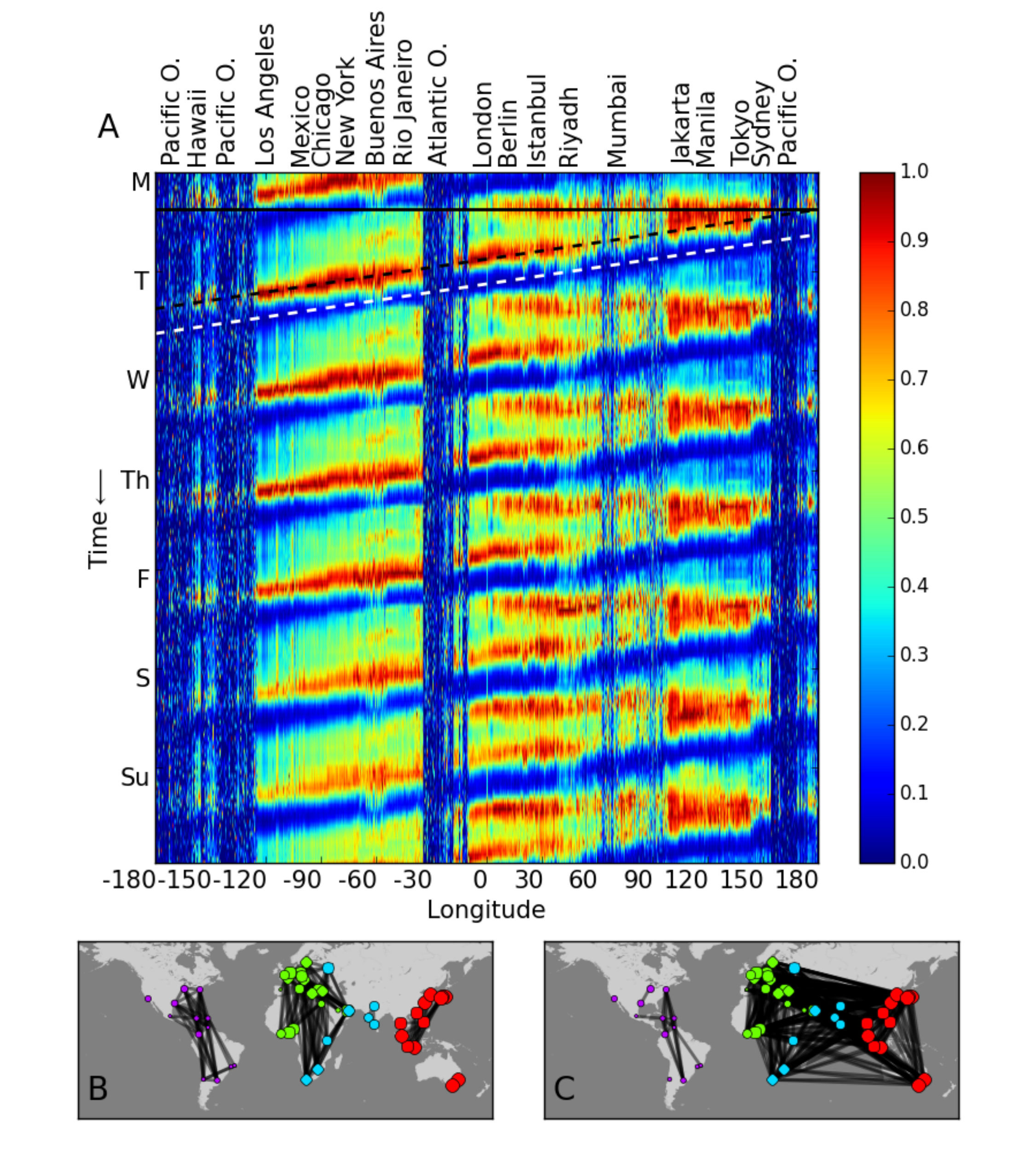}
\caption{A. Temporal dynamics of an average week of Twitter activity by longitude. The vertical axis represents time (increasing from top to bottom) and the horizontal axis represents longitude. Significant cities are indicated at the top. Diagonal dashed lines show peaks of activity (black) and inactivity (white) tracking the time of day. Horizontal line (solid black) indicates synchronous linked activity across Europe, Asia, Africa and Oceania (scale on right). B. Urban correlation network across a time window from 3pm to 3am UTC. C. Like B but across a time window from 1am to 1pm UTC. Colors indicate the results of a clustering algorithm \cite{blondel2008fast}, after aggregating the networks over time. {\color{black} An animated visualization of the urban correlation network is shown in the supplementary electronic material (video S5). }}\label{fig:WorldDynamics}%
\end{center}
\end{figure}

In Figure \ref{fig:WorldDynamics} A, we show patterns of collective behavior at a global scale after aggregating tweets by longitude. Each longitude has cycles of activity, similar to those of individual cities (Fig. \ref{fig:ActivityMap}, \ref{fig:HeartBeatCorr} and \ref{fig:UrbanNebulae}). Minima (white dashed line) and maxima (black dashed line) shift from east to west due to diurnal synchronization of sleep and wake cycles and the Earth's rotation. The ubiquity of this pattern manifests homogenization in habits and customs among globally differentiated cultures and social contexts. Furthermore, there is a specific synchronization in Figure \ref{fig:UrbanNebulae} that is actually a global phenomenon. Distinct from the other behaviors that track the daily period and therefore are diagonal, between longitudes 0 to 180, a horizontal line (black) shows a simultaneous peak of activity that occurs daily across the European, Asian, African and Oceanian continents. This horizontal peak reflects large scale dependencies across half of the world. 

The synchronization of activity is manifest in a dynamic correlation network between nodes representing cities, whose edges appear when the urban time series are correlated above a given threshold ($r>0.9$). 
{\color{black} Correlations are calculated by using overlapping time windows of 12 hours across the cities' average day time series. Highly synchronized cities have stronger connections with each other than with the rest of the network. }
In Figure \ref{fig:WorldDynamics} B and C, we show two snapshots of the correlation network at different times. Cities are generally linked within times zones, because it is natural for cities to be synchronized to those in similar longitudes (Fig. \ref{fig:WorldDynamics} B). However, at the time of global synchronization (Fig. \ref{fig:WorldDynamics} C) cities across Eurasia and Africa are strongly coupled, manifesting their synchronous activity. Such global synchronization can be expected to arise in the context of increasing global interactions consistent with synchronization in many complex systems \cite{boccaletti2002synchronization}. 

\begin{figure}
\begin{center}
\includegraphics[width=3in]{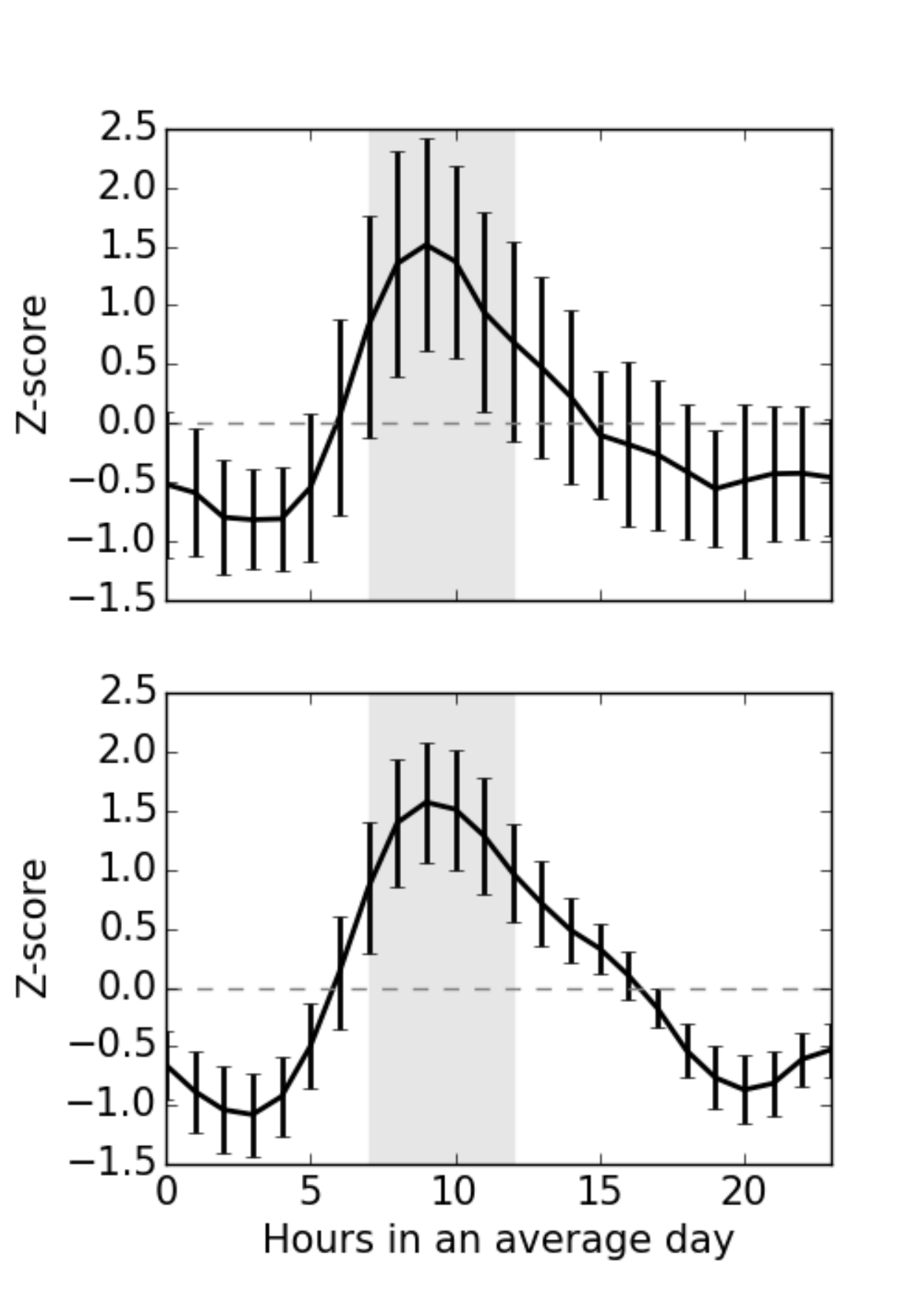}
\caption{Deviation from the average number of directed mentions (top) and shared hashtags (bottom) between the European and Asian longitude ranges during an average day in units of standard deviations (Z-score). Gray shadowed regions represent the synchronized times.}\label{fig:empirical}
\end{center}
\end{figure}

{\color{black} For additional evidence that synchronization arises from social interactions, we studied directed messages in the content of tweets (mentions) and topic identifiers (hashtags) between the European and Asian longitude ranges (see Fig. \ref{fig:empirical}). We found that the number of interurban directed messages and common hashtags simultaneously peak during the synchronized time. At this time a significantly larger number of directed messages are sent between these regions and more hashtags are shared in their messages with respect to other times of the day ($p<0.001$). These results indicate that people tend to share more information about increasingly similar topics as they synchronize their activities. }

\begin{figure}
\begin{center}
\includegraphics[width=6in]{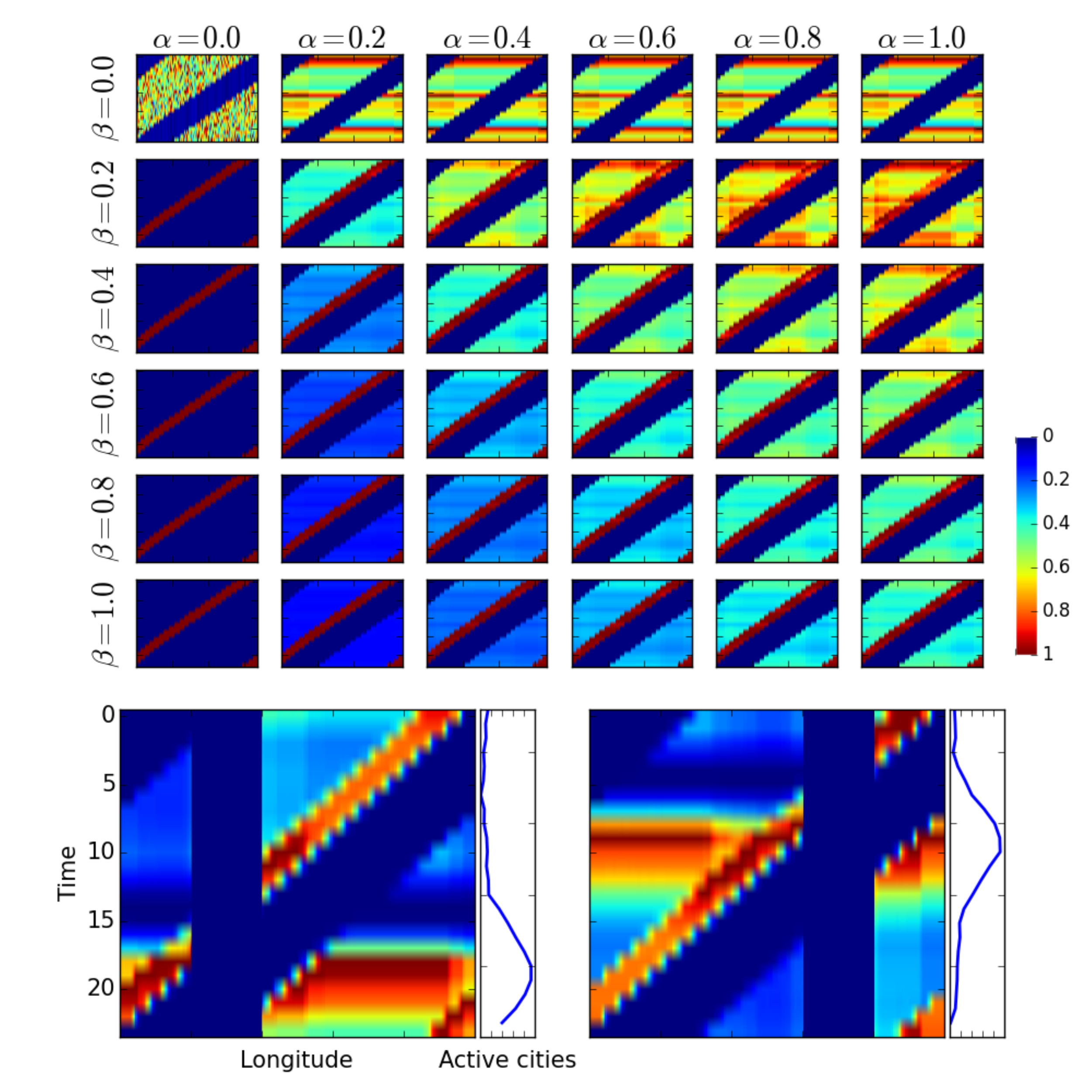}
\caption{Top panels: Model outcomes for multiple values of intercity influence at multiple longitudes ($\alpha$) and night activity ($\beta$) with the same initial conditions. Longitudes are represented in the x-axis and universal time is represented in the y-axis (see bottom left panel for scale). Colors indicate the level of activity (scale shown in figure). We normalize the cities activities by their maximum value. Bottom panels: Average model outcomes after introducing the effects of large inactive areas-oceans. Parameters $\alpha = 0.4$ and $\beta = 0.2$. The blue series indicates the number of active cities (x-axis) at each time of the day (y-axis). Color scale is similar to top panels.}\label{fig:model_force}%
\end{center}
\end{figure}

{\color{black}
We model the global synchronization process with an iterative dynamical model based on the tendency of tweets to trigger other tweets. The full model includes the tendency of people to tweet before going to bed and absence of tweets at night. The model is defined by the iterative map in the changing daily dynamics of a city: 
\begin{equation}
s_i(\sigma,\tau)=s_i(\sigma,\tau-1)+\alpha\sum_{j \ne i}{s_j(\sigma,\tau-1)}+\beta\sum_{\sigma_b}{\delta(\sigma-\sigma_b)}
\end{equation}
\noindent where $s_i$ represents the temporal activity of city $i$, $\sigma \in [0,23]$ indicates time of the day, $\tau$ represents day to day changes which is our iterative variable (we ignore weekend differences for this purpose), and $\sigma_b$ indicates the location dependent time of evening activity. Parameters $\alpha \in [0,1]$ and $\beta \in [0,1]$ indicate the weights of intercity influence at multiple longitudes and evening activity respectively. Activity is set to zero between times $0 < \sigma < 8$. We normalize $s_i(\sigma,\tau)$ at each iteration so that $\sum_\sigma s_i(\sigma,\tau) = 1$. At $\tau=0$, $s_i(\sigma)$ is set to random values.

In the top panels of Figure \ref{fig:model_force}, we present the outcomes of the model for multiple values of intercity influence ($\alpha$) and evening activity ($\beta$). The initial conditions vary among cities but do not change for different $\alpha$ and $\beta$ configurations. The blue diagonals indicate local sleeping times. If $\alpha = \beta = 0$, the model results in an independent random distribution of activities among cities. For positive values of $\alpha$, cities are able to influence each other and horizontal stripes emerge, indicating the synchronization of activities in cities at multiple longitudes. The horizontal stripes emerge at the times where the sum of initial conditions are slightly higher and therefore symmetry breaks. Similarly, higher values of $\beta$ result in peaks of activities  before sleeping time (red diagonals). Both behaviors are simultaneously found for intermediate values of these parameters (i.e. $\alpha = 0.4$ and $\beta = 0.2$).

We add an inactive ocean to the model in the bottom panel of Figure \ref{fig:model_force} and average over 100 realizations of two specific parameter values. The inactive areas affect the number of active cities at each time of the day. As a result, the synchronized peaks of activity (red horizontal stripes in squared panels) emerge at the times where most of cities are active (blue curve in rectangular panels) and consequently the sum of initial conditions is consistently higher. These results show that synchronous peaks of activity emerge from reinforcement and interaction mechanisms.
}

\section{Conclusion}

The patterns we identified manifest how social activities 
lead to dependencies among the communication between individuals, especially their synchronization. Synchronization may arise because of explicit coordination, but more generally occurs due to availability of individuals to perform actions when others trigger responses from them.
Urban synchrony is manifest in each urban area. We have found that global synchrony appears to be arising in the context of increasing global communication. 
Through synchronization, the timing of human communication and activities becomes constrained to the norms and conventions of the social environment.
As people create and maintain temporally sensitive relationships, the complexity of the larger scale behavior of the social system increases. These collective behaviors may be linked to collective capabilities that provide products, services and information that cannot be obtained by an individual alone.

\section{Author Contributions} 
All authors contributed in making the figures and writing the manuscript.

\section{ Acknowledgments} 
We thank Richard J. Cohen for helpful comments and discussions.

\section{ Funding Statement} 
NECSI internal funds.


\end{document}